\documentclass[referee]{aa520}
\usepackage{graphicx}
\usepackage{psfig}
\usepackage{epsfig}
\begin{document}
\title{Photometry and low resolution spectroscopy of hot post-AGB candidates}
\author{G. Gauba \inst{1}
\and M. Parthasarathy \inst{1}
\and Brijesh Kumar \inst{2}
\and R.K.S. Yadav \inst{2}
\and \\Ram Sagar \inst{1,2}}
\institute{Indian Institute of Astrophysics, Koramangala, Bangalore 560034,
India
\and State Observatory, Manora Peak, Nainital 263129 (Uttaranchal), India}
\offprints{G. Gauba,
\email{gauba@iiap.ernet.in}}
\date{Received / Accepted}
\authorrunning{G.Gauba et al.}
\titlerunning{Photometry and spectroscopy of hot post-AGB candidates}

\abstract{
We have obtained Johnson U, B, V and Cousins R, I photometry 
and low resolution spectra of a small sample of hot post-AGB 
candidates. Using the present data in combination with JHK data 
from 2MASS, infrared data from the MSX catalog and the 
IRAS fluxes, we have studied the spectral energy 
distribution (SED) of these stars. Using the DUSTY code we have
estimated the dust temperatures, the distances to the stars, 
the mass-loss rates, angular radii of the inner boundary of
the dust envelopes and dynamical ages from the tip of the AGB.
These candidates have also been imaged through 
a narrow band H$\alpha$ filter, to search for nebulosity around the 
central stars. Our H$\alpha$ images revealed the bipolar morphology 
of the low excitation PN IRAS 17395-0841 with an angular extent of 
2.8\arcsec. The bipolar lobes of IRAS 17423-1755 in H$\alpha$ were 
found to have an angular extent of 3.5\arcsec (south-east lobe) and 
2.2\arcsec (north-west lobe). The dust envelope characteristics, 
low resolution spectrum and IRAS colors suggest that IRAS 18313-1738 is  
similar to the proto-planetary nebula (PPN) HD 51585. 
The SED of IRAS 17423-1755, IRAS 18313-1738 and IRAS 19127+1717 show a
warm dust component (in addition to the cold dust) which may 
be due to recent and ongoing mass-loss.
}
\maketitle

\section{Introduction}

Stars having initial mass between 0.8 and 8 M$_{\odot}$ pass
through the asymptotic giant branch (AGB) phase of evolution 
(Iben \& Renzini, 1983). These low and intermediate mass stars, 
undergo severe mass-loss (10$^{-7}$ - 10$^{-4}$ M$_{\odot}$ yr$^{-1}$) 
on the AGB. The post-AGB phase is the rapid evolution of the star
from the tip of the AGB to the planetary nebula (PN) stage.
During the post-AGB phase the circumstellar dust shell moves away from 
the star and decreases in temperature from $\sim$ 400 to 100K 
(Bedijn, 1987; Zijlstra et al., 1992). The rapidly evolving, 
hotter analogues (OB spectral types) of the cooler (G,F,A) post-AGB stars 
(Parthasarathy \& Pottasch 1986, 1989) are interesting to study 
in order to understand the mass-loss mechanisms, 
dust envelope characteristics and dynamical time scales responsible for the 
evolution of low and intermediate mass stars to PNe of varied morphologies. 
In this paper, we analyse the complete spectral energy distribution of a 
small sample of hot post-AGB candidates. We have also searched for
nebulosity around these stars. As part of our search for new 
hot post-AGB candidates, we also present the spectra of two high galactic 
latitude early B-type supergiants.

\section{Target Selection}

Hot post-AGB stars may be identified on the basis of their
high temperature, low surface gravity, evidence of remnant
AGB envelope (infrared excess), detached dust envelope, double
peaked spectral energy distribution and nebulosity due to 
scattered light (Parthasarathy et al., 2000a, Kwok, 2001). 
Pottasch et al. (1988) and van der Veen \& Habing (1988) identified
a region of the IRAS color-color diagram (F(12$\mu$)/F(25$\mu$) $<$ 0.35
and F(25$\mu$)/F(60$\mu$) $>$ 0.3) which was mainly populated
by stars in transition from the AGB to the PN phase. An occasional
HII region, Seyfert galaxy, or T-Tau star is not excluded from this range
(Pottasch et al., 1988, Szczerba et al., 2001).
But in the case of post-AGB stars, the circumstellar dust temperatures 
are in the range of 100 - 200K (Kwok, 2001). In addition to the 
cold dust, warm dust indicative of recent or ongoing
post-AGB mass loss may also be present in the circumstellar envelopes
of these stars (eg. Hen401, Parthasarathy et al., 2001).
Young massive OB supergiants are not expected at high galactic latitudes. Also
young massive OB supergiants do not have detached cold circumstellar dust
shells. High galactic latitude OB supergiants with detached dust shells
and far-infrared colors similar to PNe were indeed found to be in the
post-AGB stage of evolution (Parthasarathy, 1993a, Parthasarathy et al., 2000a).
To firmly establish the evolutionary status it is important to obtain
high resolution optical spectra of the candidate stars for determination 
of their chemical abundances and to look for signatures of AGB 
nucleosynthesis.

The targets selected for photometry (Table 1) were IRAS sources with OB
spectral types having F(12$\mu$)/F(25$\mu$) $<$ 0.35 and 
F(25$\mu$)/F(60$\mu$) $>$ 0.3. Based on their low resolution spectra and
spectral type, many of these targets had been classified as post-AGB stars by 
Parthasarathy et al. (2000a). Drilling \& Bergeron (1995) extended the 
survey of luminous stars in the Milky Way to galactic latitude  
b = $\pm$ 30$^{\circ}$ for l = $\pm$ 60$^{\circ}$. They designated this 
as the LSE survey. We have also selected few LSE stars at high 
galactic latitudes with OB-spectral types (Table 2). 
The optical counterparts of the IRAS sources were carefully 
identified using the Digitised Sky Survey and NASA's Sky View 
Java interface. The targets were searched within the IRAS error boxes. 
Besides most of the targets are in scarcely populated regions at 
high galactic latitudes. 

\begin{table}
\begin{center}
\caption{List of hot post-AGB candidates selected for photometry}
\begin{tabular}{|c|c|c|c|c|c|c|c|c|}
\hline
IRAS&Name&l&b&Sp.Type& 
\multicolumn{4}{c|}{IRAS Fluxes (Jansky)} \\
     &      &   &   &          &        
12$\mu$ & 25$\mu$ & 60$\mu$ & 100$\mu$ \\ 
\hline \hline

17074-1845& Hen 3-1347 &  4.1 & +12.3 & B3IIIe &  
0.50 & 12.20 & 5.66 & 3.47: \\

17395-0841& SS 318     & 17.0 & +11.1 & PN &  
0.31 &  4.18 & 8.43 & 6.38 \\

17423-1755& Hen 3-1475 &  9.4 &  +5.8 & Be & 
7.05 & 28.31 & 63.68 & 33.43 \\

18237-0715& MWC 930    & 23.6 &  +2.2 & Be &  
1.84 & 4.01 & 38.24 & 35.13 \\

18313-1738& MWC 939    & 15.3 &  -4.3 & Be & 
9.41 & 7.28 & 1.00: & 68.70L \\

19127+1717& SS 438     & 51.0 &  +2.8 & B9V & 
12.15 & 18.79 & 8.50 & 7.37L \\

19157-0247& LS IV -0229& 33.6 &  -7.2 & B1III & 
8.88 & 7.16 & 2.45 & 11.05L \\

19200+3457& StHA161    & 67.6 &  +9.5 & B &
0.25L & 2.12 & 1.45 & 1.41L \\

19399+2312& LS II +2317& 59.3  &  +0.1 & B1III & 
1.13 & 2.27: & 23.31L & 80.20L \\

\hline
\end{tabular}

\vspace{0.2cm}

\noindent \parbox{16cm}{A colon {\bf :} indicates moderate quality IRAS flux, 
{\bf L} is for an upper limit}
\end{center}
\end{table}

\begin{table}
\begin{center}
\caption{Characteristics of the hot post-AGB candidates selected 
spectroscopically}
\begin{tabular}{|c|c|c|c|c|c|c|c|c|}
\hline
Object & R.A. & DEC & Sp. & V & E(B$-$V) & l & b & d (kpc) \\
   & (2000) & (2000)& Type&   &          &   &   &         \\     
\hline \hline

LSE 163= & 13:08:46 & -43:27:51 & B2 I & 10.4  & 0.1 & 
306.3 & +19.3 & 4.3 \\
CD -42 8141 &       &           &      &       &     &
      &       & \\
\hline
LSE 45=  & 13:49:18 & -50:22:46 & B2 I & 11.0  & 0.2 & 
312.3 & +11.4 & 4.8 \\ 
CD -49 8217 &       &           &      &       &     &
      &       &  \\ 
\hline

\end{tabular}
\end{center}
\end{table}

\section{Observations}

U,B,V,R,I images of the hot post-AGB candidates were obtained on
10 April, 2000 at the State Observatory (SO) with the f/13 
Cassegrain 104 cm. Sampurnanand Carl-Zeiss telescope. The central 
1280$X$1280 pixels of a 2K$X$2K CCD were used for the purpose, 
in 2$X$2 binning mode, resulting in 680$X$680 pixels image frames. 
The PG 1323-086 standard stars field (Landolt, 1992) was observed 
several times during the night for calibration and extinction 
measurements. 

The narrow band images of the stars were obtained with the same 
telescope on 11 April, 2000 with a 1K$X$1K CCD having plate
scale of 0.37\arcsec pixel$^{-1}$. An H$\alpha$ filter centered 
at 6565\AA~ with a bandwith of 80\AA~ and a continuum filter 
centered at 6650\AA~ with a bandwith of 80\AA~ were used for 
the purpose. 

Optical spectra of the hot post-AGB candidates were obtained with 
the 1.02m telescope at the Vainu Bappu Observatory (VBO) on 20 January,
2000 with a resolution of 1.1\AA~ pixel$^{-1}$. Appropriate number of 
flat fields and bias frames were observed on 
each night. Following each observation of a program star, we 
obtained a comparison lamp spectra and the spectra of a bright 
star close to the RA \& DEC of the object. 

\setcounter{table}{2}
\begin{table}
\begin{center}
\renewcommand{\thetable}{\arabic{table}a}
\caption{Photometry of hot post-AGB candidates}
\begin{tabular}{|c|c|c|c|c|c|c|c|c|c|}
\hline
IRAS & U & B & V & R & I & J & H & K & H$_{\alpha}$ \\
name &mag&mag&mag&mag&mag&mag&mag&mag& emission \\
\hline

17074-1845 &  --    & 11.92$^{1}$ & 11.55 & 11.25 & 11.07 &11.33$^{2}\pm$ &12.06$^{2}\pm$ &
11.02$^{2}\pm$ &NO \\
           &        &        &        &        &        & 0.13 & 0.16 & 0.11 & \\   
\hline
17395-0841 & 15.37 & 15.04 & 13.74 & 12.79 & 11.94 &10.59$^{2}\pm$ & 9.76$^{2}\pm$ & 9.21$^{2}\pm$ &YES\\
           &        &        &        &        &        & 0.06      & 0.03      &0.04 & \\ 
\hline
17423-1755 & 13.57 & 13.3$^{3}$ & 12.64 & 11.75 & 10.91 & 9.61$^{2}\pm$ & 8.32$^{2}\pm$ & 6.80$^{2}\pm$ &YES\\
           &        &        &        &        &        & 0.05      & 0.02      &0.02 & \\      
\hline
18237-0715 & 15.48 & 14.88 & 12.37 & 10.51 &  8.78 & -- & -- & -- & YES\\
\hline
18313-1738 & 12.36 & 12.88 & 12.37 & 11.60 & 11.37 &9.995$\pm$ &8.374$\pm$ &6.785$\pm$ &YES\\
           &        &        &        &        &        & 0.027 & 0.045 & 0.019 & \\      
\hline
19127+1717 & 14.70  & 14.20  & 13.13 & 12.38  &  8.65 & 11.07$^{4}$ & 
9.94$^{4}$ & 8.65$^{4}$ & NO\\
\hline
19157-0247 & 11.62 & 11.31 & 10.55 & 10.04 &  9.48 & -- & -- & -- & NO\\
\hline
19200+3457 & 10.59 & 11.35  & 11.25  & 11.14 & 11.09 & 11.008$\pm$
& 10.883$\pm$ & 10.739$\pm$ & NO\\
           &        &        &        &        &        &  0.024 &  0.025 &  0.032 & \\       
\hline
19399+2312 &  --   & 10.90 & 10.05 &  9.46 &  8.85 &  8.573$\pm$ &  
8.388$\pm$ & 8.331$\pm$ & NO\\
           &        &        &        &        &        &  0.022 &  0.003 & 0.033 & \\

\hline
\end{tabular}\\
\vspace{0.7cm}
\noindent \parbox{16cm}{$^{1}$ The B magnitude of IRAS17074-1845 is 
from the Tycho-2 Catalog (Hog et al., 2000)\\ 
$^{2}$ From Garcia-Lar\'io, Manchado, Pych, \& Pottasch (1997) \\
$^{3}$ The B magnitude of IRAS17423-1755 is from the USNO-A2.0 Catalog
(Monet et al., 1998)\\
$^{4}$ From Whitelock \& Menzies (1986)\\
$^{5}$ The rms errors in U,B,V,R,I obtained from fitting the 
transformation equations to the standard stars\\
are $\pm$ 0.05, $\pm$ 0.05, $\pm$ 0.05, $\pm$ 0.09, $\pm$ 0.07 respectively}
\end{center}
\end{table}

\setcounter{table}{2}
\begin{table}
\begin{center}
\renewcommand{\thetable}{\arabic{table}b}
\caption{MSX data}
\vspace{0.2cm}
\begin{tabular}{|c|c|c|c|c|}
\hline
&\multicolumn{4}{c|}{MSX Fluxes (Jansky)}\\ \cline{2-5}
IRAS   & Band A & Band C & Band D & Band E \\
    & 8.28$\mu$ & 12.13$\mu$ & 14.65$\mu$ & 21.34$\mu$ \\
\hline \hline
18313-1738 & 8.9623 & 7.5263 & 6.0883 & 8.3414 \\
19127+1717 & 8.332  & 11.737 & 12.577 & 16.636 \\

\hline
\end{tabular}
\end{center}
\end{table}

\begin{table}
\begin{center}
\caption{Input physical parameters for DUSTY and the adopted reddening
values}
\begin{tabular}{|c|c|c|c|c|c|c|}
\hline
IRAS & dust & E(B$-$V) & E(B$-$V) & T$_{*}$ & grain & Optical \\
     & type & C.S.+I.S.&  I.S.    &   (K)   & type  & depth ($\tau$)\\                        
\hline \hline

17074-1845 & cold & 0.66 & 0.28 & 17100 & Sil-DL & 0.105 \\ \hline
17395-0841 & cold & 1.60 & 1.23 & 35000 & amC-Hn & 0.002 \\ \hline
17423-1755 & warm & 0.86 & 0.67 & 20000 & amC-Hn & 0.16  \\  
           & cold & 0.86 & 0.67 & 20000 & SiC-Pg & 0.35  \\ \hline
18237-0715 & cold & 2.71 &  --  & 20000 & SiC-Pg & 0.004 \\ \hline
18313-1738 & warm & 0.71 &  --  & 20000 & grf-DL & 0.2   \\ 
           & cold & 0.71 &  --  & 20000 & Sil-DL & 0.16  \\ \hline 
19127+1717 & warm & 1.14 &  --  & 10500 & grf-DL & 0.07  \\
           & cold & 1.14 &  --  & 10500 & Sil-DL, grf-DL \& amC-Hn & 0.7 \\ \hline 
19157-0247 & cold & 1.02 & 0.65 & 24000 & SiC-Pg & 0.004 \\ \hline
19200+3457 & cold & 0.30 & 0.14 & 20000 & Sil-Ow & 0.04 \\ \hline 
19399+2312 & cold & 1.11 &  --  & 24000 & SiC-Pg & 0.006 \\ \hline

\end{tabular}
\end{center}
\end{table}

\setcounter{table}{4}
\begin{table}
\begin{center}
\renewcommand{\thetable}{\arabic{table}a}
\caption{Derived stellar and dust envelope parameters for M$_{c}$=0.565}
\begin{tabular}{|c|c|c|c|c|c|c|c|c|}
\hline
IRAS & T$_{d}$ & r1   & r0 & d     & $\theta$  & $\dot M$          & V$_{e}$ & $\Delta$t \\ 
     &   (K)   & (cm) & (cm) & (kpc) & (\arcsec) & M$_{\odot}$yr$^{-1}$ & kms$^{-1}$ & yr. \\ 
\hline \hline
17074-1845&122&3.9X10$^{16}$& 1.1X10$^{14}$ & 3.1 & 0.8 & 1.0X10$^{-5}$& 17.4&743\\ \hline
17395-0841&110&1.7X10$^{17}$& 1.4X10$^{14}$ & 1.0 & 11.7& 1.1X10$^{-6}$& 7.7 &7344 \\ \hline
17423-1755&100&2.9X10$^{17}$& 1.5X10$^{14}$ & 3.1 &  6.3& 5.3X10$^{-5}$& 23.3&4154\\ \hline
18237-0715&100&1.4X10$^{17}$& --   & 0.2 & 45.0 &   --          &-- &-- \\ \hline
18313-1738&450&2.3X10$^{15}$& 1.1X10$^{14}$ & 3.6 & 0.04& 3.6X10$^{-6}$& 18.9&39\\ \hline
19127+1717&300&6.2X10$^{15}$& 1.4X10$^{14}$ & 4.0 &  0.1& 9.0X10$^{-6}$& 25.9&78\\ \hline
19157-0247&135&1.1X10$^{17}$& 1.4X10$^{14}$ & 0.8 &  9.1& 1.4X10$^{-6}$& 8.8&4180\\ \hline
19200+3457&140&3.1X10$^{16}$& 1.5X10$^{14}$ & 3.8 &  0.5& 4.0X10$^{-6}$& 12.8&805\\ \hline
19399+2312&127&1.3X10$^{17}$& --   & 0.6 & 16.0 &   --          &-- &-- \\ \hline

\hline
\end{tabular}\\
\end{center}
\end{table}

\setcounter{table}{4}
\begin{table}
\begin{center}
\renewcommand{\thetable}{\arabic{table}b}
\caption{Derived stellar and dust envelope parameters for M$_{c}$=0.605}
\begin{tabular}{|c|c|c|c|c|c|c|c|c|}
\hline
IRAS & T$_{d}$ & r1   & r0 & d     & $\theta$  & $\dot M$      & V$_{e}$  & $\Delta$t \\
     &   (K)   & (cm) & (cm) & (kpc) & (\arcsec) & M$_{\odot}$yr$^{-1}$ & kms$^{-1}$ &  yr. \\
\hline \hline
17074-1845&122&4.6X10$^{16}$ & 1.8X10$^{14}$ & 3.7 &  0.8&1.3X10$^{-5}$&20.6&743\\ \hline
17395-0841&110&2.0X10$^{17}$ & 2.3X10$^{14}$ & 1.1 & 11.8&1.5X10$^{-6}$&8.8&7543\\ \hline
17423-1755&100&3.5X10$^{17}$ & 2.6X10$^{14}$ & 3.7 &  6.3&6.8X10$^{-5}$&26.0&4481\\ \hline 
18237-0715&100&1.7X10$^{17}$ & --   & 0.2 & 45.0 &    --        & -- &--\\ \hline
18313-1738&450&2.7X10$^{15}$ & 1.9X10$^{14}$ & 4.2 & 0.04&4.7X10$^{-6}$&22.3&38\\ \hline
19127+1717&300&7.4X10$^{15}$ & 2.3X10$^{14}$ & 4.7 &  0.1&1.2X10$^{-5}$&28.7&84\\ \hline
19157-0247&135&1.3X10$^{17}$ & 2.4X10$^{14}$ & 0.9 &  9.2&1.8X10$^{-6}$&9.9&4336\\ \hline
19200+3457&140&3.7X10$^{16}$ & 2.5X10$^{14}$ & 4.6 &  0.5&5.1X10$^{-6}$&14.7&834\\ \hline
19399+2312&127&1.6X10$^{17}$ & --   & 0.7 & 16.0 &     --      & -- &--\\ \hline

\end{tabular}\\
\end{center}
\end{table}

\section{Analysis of photometric and spectroscopic data}

The DAOPHOT package in IRAF was used for the reduction and analysis of
the photometric data (Massey \& Davis, 1992). Bias subtraction and flat 
fielding was performed on the individual image frames. The extinction 
coefficient for each filter was determined from the standard star 
observations. The observed photometric magnitudes of the objects are 
listed in Table 3a. The 2MASS (2Micron All Sky Survey) Catalog was searched 
within 15\arcsec of each object for their JHK magnitudes. Whenever possible, 
the JHK magnitudes have been taken from Garc\'ia-Lario et al. (1997). We 
also searched the MSX (Midcourse Space Experiment) catalog within 3\arcsec 
of the objects. The MSX fluxes are listed in Table 3b. The spectral types of 
the objects have been taken from literature. 

The objects were imaged through the H$\alpha$ and continuum filters.
To detect H$\alpha$ emission, we followed the procedure outlined by 
Beaulieu et al. (1999). The continuum images were scaled to match the 
total star signal above the sky in their corresponding H$\alpha$ 
images. An appropriate offset was then applied to each continuum image 
to equate the sky level to that of the H$\alpha$ image. To find the 
H$\alpha$ emitting objects, we then divided the H$\alpha$ image by the 
continuum image for each field. We detected H$\alpha$ emission in four
of the objects. Fig. 1 shows the continuum and resultant H$\alpha$ images
of the objects. 

\begin{figure}
\psfig{figure=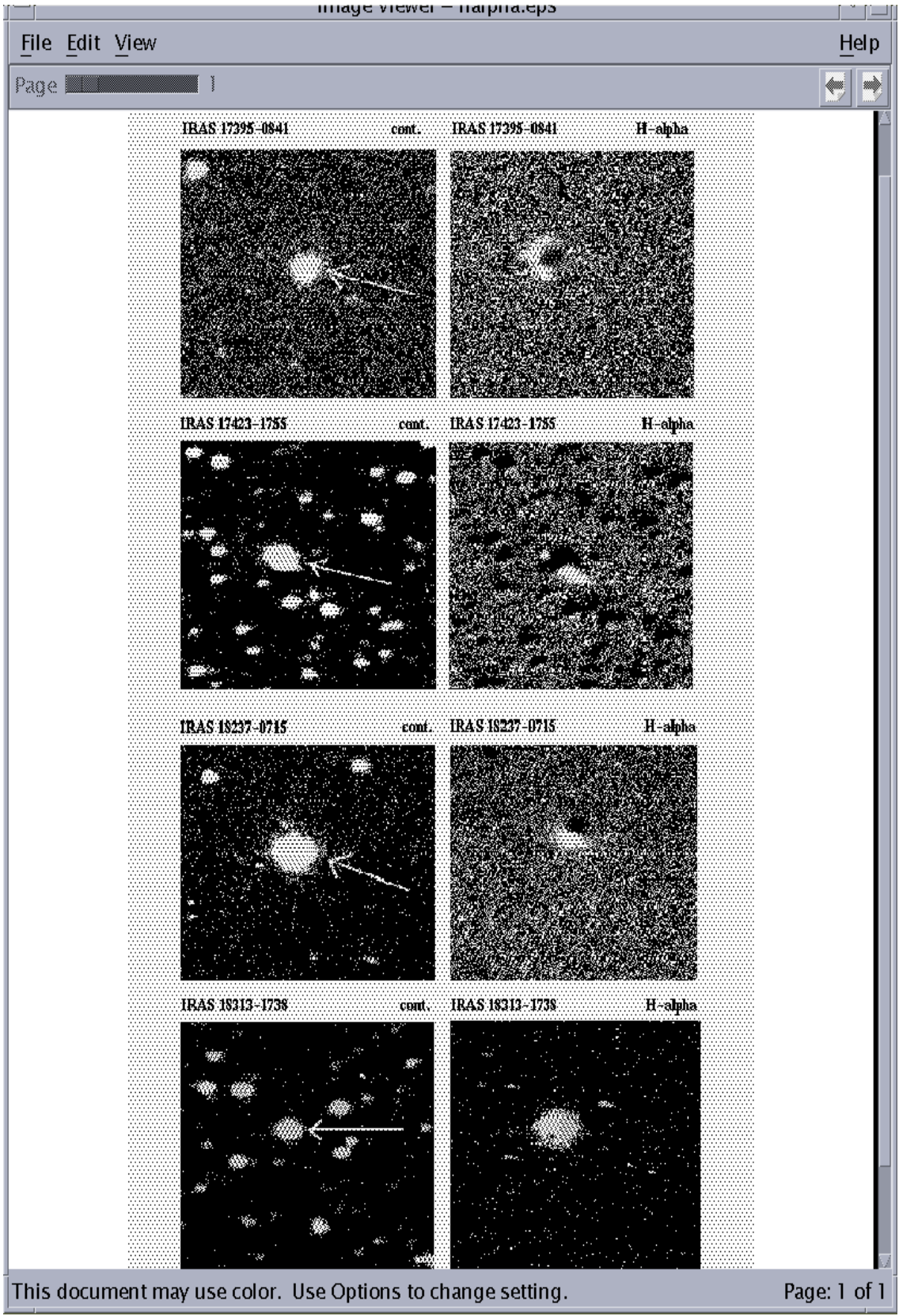,height=22cm,width=15cm}
\caption{Continuum ( $\lambda$=6650\AA~, $\Delta\lambda$=80\AA~) and 
H$\alpha$ ( $\lambda$=6565\AA~, $\Delta\lambda$=80\AA~) 
images of the selected hot post-AGB stars}
\end{figure}

The optical spectra were bias subtracted, flat field corrected, 
wavelength calibrated and continuum normalised using the IRAF package. 
Fig.2 shows the continuum normalised spectra from 4250\AA~ to 5500\AA~.
Spectral classification was performed using the spectra of standard stars 
from Jacoby et al. (1984).

\begin{figure}
\epsfig{figure=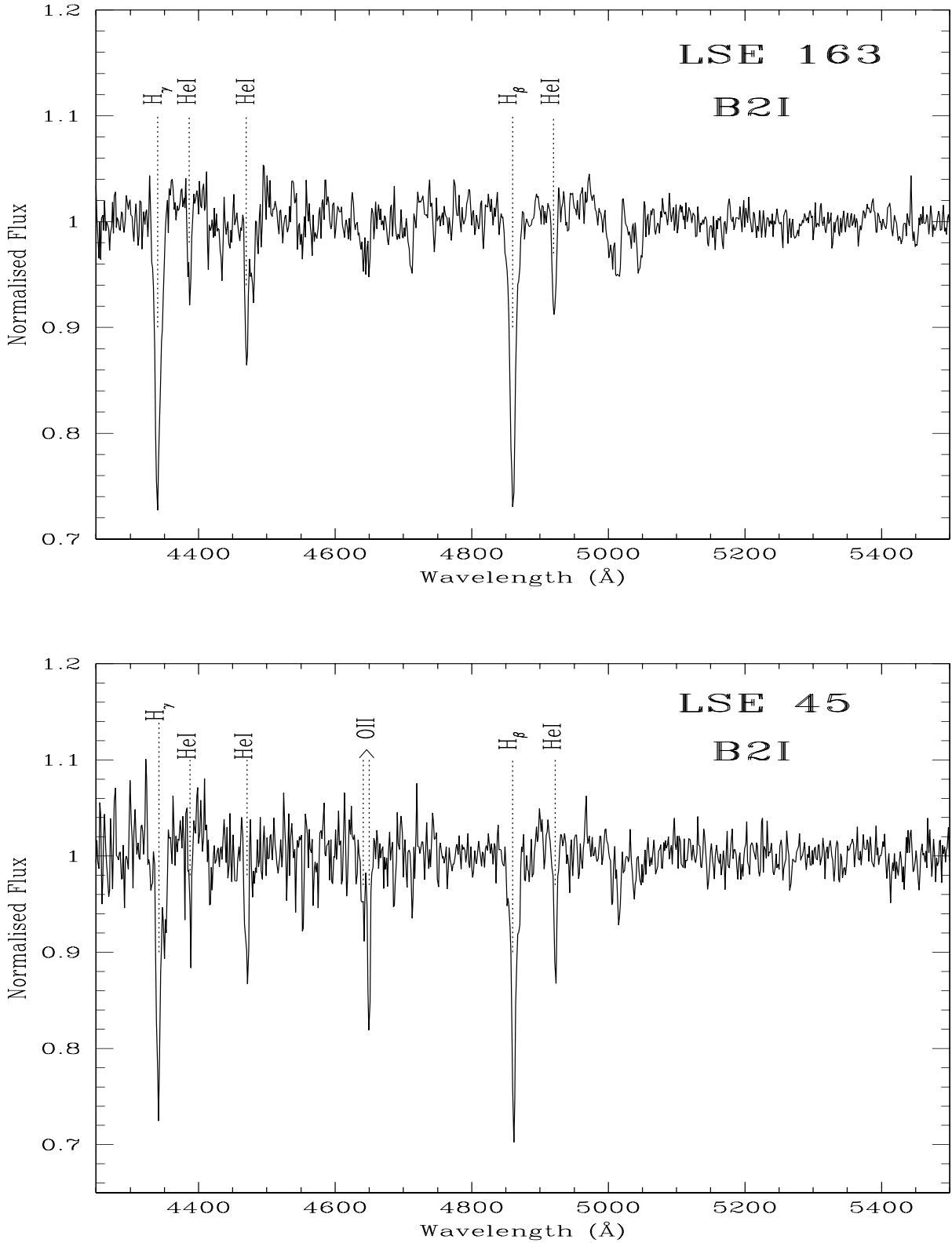,height=21cm,width=16cm}
\caption{Low resolution spectra of LSE 163 and LSE 45}
\end{figure}

\subsection{Central star temperatures and reddening}

Whenever, the spectral type and luminosity class of the star was 
known, the effective temperature of the central star, was taken
from Lang(1992). For Be stars and for the bipolar PPN Hen3-1475, we
assumed a central star temperature of 20000K. The total extinction
(I.S.+C.S.) towards the stars due to interstellar and circumstellar dust grains
was estimated from the difference between the observed and intrinsic
(B-V) values. For Be stars, we assumed intrinsic (B$-$V) = $-$0.20 
corresponding to an effective temperature of 20000K and for the low 
excitation PN IRAS 17395-0841 we assumed intrinsic (B$-$V) = $-$0.3 
corresponding to an effective temperature of 35000 K for the hot 
central star. Since the distances to the stars were not known, the 
interstellar extinction (I.S.) in the direction of the stars was estimated 
using the Diffuse Infrared Background Experiment (DIRBE)/IRAS dust maps
(Schlegel et al., 1998). The DIRBE/IRAS reddening estimates have
an accuracy of 16\%. The difference between the two sets of values would give 
an estimate of the extinction due to circumstellar dust. The DIRBE/IRAS 
dust maps do not give reliable estimates of the interstellar extinction
for $|$b$|$ $<$ 5$^{\circ}$. The U,B,V,R,I and near infrared magnitudes 
of the stars were corrected for the total extinction 
(E(B$-$V) I.S.+C.S.) assuming R$_{v}$ = 3.1. 

\subsection{Modelling of the circumstellar dust envelopes with DUSTY code} 

To derive physical parameters from the spectral energy distribution
of the stars, we solved the problem of radiation transfer through 
a dust envelope using the DUSTY code (Ivezi\'c et al., 1999) developed at the 
University of Kentucky assuming centrally-heated spherical density
distributions. The central stars were assumed to be point sources, at the center of 
the spherical density distributions and their spectral energy distributions
were taken to be Planckian. For modelling, we chose from six different grain types :
'warm' (Sil-Ow) and 'cold' (Sil-Oc) silicates from Ossenkopff et al. (1992), 
silicates and graphites (Sil-Dl and grf-DL) from Draine and Lee (1984), 
amorphous carbon (amC-Hn) from Hanner (1988) and SiC (SiC-Pg) from P\'egouri\'e (1988).
DUSTY contains data for the optical properties of these six grain types. 
For all stars except IRAS17395-0841 and MWC930
the standard Mathis, Rumpl, Nordsieck (MRN) (Mathis et al., 1977) 
power-law was used for 
the grain size (n(a)) distributions, i.e. 
n(a) $\propto$ a$^{-q}$ for a$_{min}$ $\le$ a $\le$ a$_{max}$
with q=$-$3.5, a(min)=0.005$\mu$ and a(max)=0.25$\mu$. For 
IRAS17395-0841 and MWC930, we used the MRN distribution with
q=4.3, a(min)=0.005$\mu$, a(max)=0.25$\mu$ and q=3.7, a(min)=0.10$\mu$
and a(max)=0.25$\mu$ respectively. The dust temperature (T$_{d}$) on the 
inner shell boundary and the
optical depth ($\tau$) at 0.55$\mu$ were varied assuming an inverse square law (y$^{-2}$)
for the spherical density distribution. The shell was assumed to extend to 1000 times
its inner radius. For MWC930 and LSII+2317, we had to adopt a different 
power law (y$^{-p}$) for the density distribution in the shell with p=1.3 and 0.55 
respectively inorder to obtain a good fit. DUSTY does not allow simultaneous
modelling of warm and cold dust shells. Hence, the cold dust in the case of
Hen3-1475, MWC939 and SS438 had to be modelled and treated independent of the
warm dust in these stars. We adopted the fits for which the sum of squares 
of the deviations between the observed and modelled fluxes (after scaling)
were a minimum.  Table 4 lists the adopted input parameters. Fig. 3 shows
the spectral energy distribution of the stars.

Having fixed T$_{d}$ and $\tau$, we then used the gas-dynamical 
mode of the DUSTY code, to derive the mass-loss rate
for a given model. However, in the case of MWC930 and LSII+2317 where the 
density distribution follows a different law, a hydrodynamics calculation and hence
the determination of mass-loss rate was not possible. The output of the code
gives the shell inner radius, r1(cm) where the dust temperature (T$_{d}$) is 
specified. The radius scales in proportion to L$^{1/2}$ where L is the luminosity
and the output value corresponds to L=10$^{4}$L$_{\odot}$. The mass-loss rate
($\dot M$) scales in proportion to L$^{3/4}$(r$_{gd}\rho_{s})^{1/2}$ where,
the gas-to-dust mass ratio, r$_{gd}$=200 and the dust grain density, 
$\rho_{s}$=3 g cm$^{-3}$. We carried out calculations for hot post-AGB stars 
with core masses of 0.565M$_{\odot}$ and 0.605M$_{\odot}$ corresponding 
to luminosities of 4500L$_{\odot}$ (Sch\"onberner, 1983) and 
6300L$_{\odot}$ (Bl\"ocker, 1995) respectively. The distances (d) were derived 
using r1 and the ratio of the observed and
modelled fluxes at 0.55$\mu$. 

$\theta$ (=r1/d) is the angular radii of the inner boundary of the
cold circumstellar dust envelopes. In some cases eg. IRAS19399+2312,
$\theta$ is very large (16\arcsec) but the nebula is not detected
in the optical and in the case of IRAS18237-0715, $\theta$ = 45\arcsec
~but the detected nebulosity is less than 2\arcsec. In these cases, the
scattered light radiated by the dust shells in the optical may be 
too faint to be seen by our observations. In the V-band, some of our stars
have a flux of the order of 6.9X10$^{-11}$ erg s$^{-1}$ cm$^{-2}$,
eg. IRAS 17395-0841 (V=13.74). The reflection nebular (scattered light) 
flux per pixel on the CCD frames will be several orders of magnitude 
fainter. Our maximum exposure in H$\alpha$ was 5 minutes which is not
sufficient enough to reach our telescope's detection limit in H$\alpha$.
As we move away from the central star, the scattered light flux drops 
down rapidly (see eg. the radial intensity profile of AFGL 2688 in 
Kwok et al., 2001). Longer exposures would be required to detect the fainter
reflection nebulae. High resolution HST images (eg. Su et al., 2001) and
subarcsecond mid-infrared imaging (Kwok et al., 2002) have been
used to resolve PPNe.

The time when the star leaves the AGB can be assumed to coincide with
the ejection of its dust envelope. For post-AGB stars with core masses
of 0.565 M$_{\odot}$ and 0.605 M$_{\odot}$ the effective temperatures
at the tip of the AGB are 5000K (Sch\"onberner, 1983) and 6000K (Bl\"ocker, 1995)
respectively. The dust condensation temperature is taken to be 
1200K (Whittet, 2003). Using the hydrodynamic mode of the DUSTY code,
we then obtained the terminal outflow velocities (V$_{e}$) and the distances
(r0) at which the dust grains reached the condensation temperature. Ve 
scales in proportion to  L$^{1/4}$(r$_{gd}\rho_{s})^{-1/2}$. The present 
inner boundary of the dust envelopes is at r1. Thus, the dynamical age 
from the tip of the AGB 
is given by (r1$-$r0)/V$_{e}$.  The V$_{e}$ values are in good agreement 
with the expansion velocity of 10 kms$^{-1}$ assumed by 
Volk and Kwok (1989) for models of PPNe and the expansion velocities
for PPNe estimated from CO measurements by Loup et al. (1990). There
is an inherent uncertainity of 30 $\%$ in the ($\dot M$) and V$_{e}$
values obtained with the DUSTY code (Ivezi\'c et al., 1999). 

Tables 5 a and b list the respective values for T$_{d}$, r1, r0, d,  
$\theta$, $\dot M$, V$_{e}$ and $\Delta$t. All calculations were carried out 
using the best fit parameters for the cold circumstellar dust shells.

\subsection{The LSE stars}

On comparing the low resolution spectra of the high galactic 
latitude stars, LSE163 and LSE45 with that of standard stars 
from Jacoby et al. (1984) we classified them as B2I. They may be similar 
to PG 1323-086 and PG 1704+222, two high galactic latitude B-type 
post-AGB stars (Moehler \& Heber, 1998). Using the relation
between core-mass and quiescent luminosity maximum for AGB stars 
(Wood \& Zaro, 1981), with a typical core-mass of 0.6 M$_{\odot}$, we 
estimated a value of 10$^{3.79}$ L$_{\odot}$ for the luminosity of the 
central stars. Since, M$_{bol}$(Sun) = 4.75, we obtained an absolute bolometric 
magnitude of $-$4.73 for the stars. Applying the bolometric correction for the
spectral type of the star (Lang,1992), and using the distance modulus
method, we obtained distance estimates of 4.3 and 4.8 kpc respectively to these stars. 
If on the other hand, we adopt the absolute visual magnitude
(M$_{v}$) of -6.4 for a normal Population I B-type supergiant we obtain
distance estimates of 19.3 and 21.7 kpc respectively. Such large
distance estimates, suggesting that these objects
lie at the outer edge of our galaxy appear to be unphysical. Abundance
analysis from the high resolution spectra of these objects may
confirm their evolutionary status as evolved low mass post-AGB stars.

\begin{figure}
\epsfig{figure=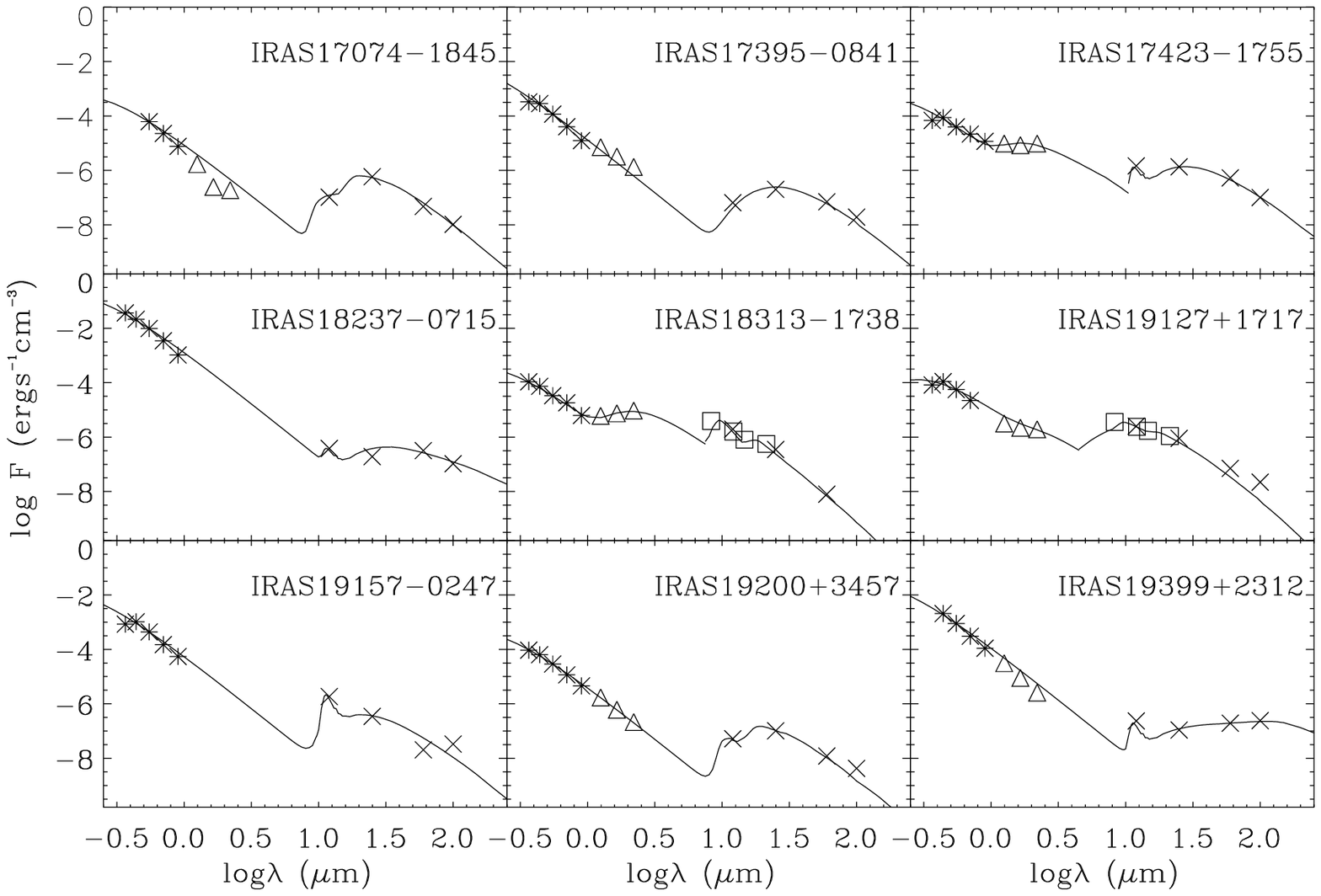,height=20cm,width=18.5cm}
\caption{Spectral Energy distributions (SED) of the 
hot post-AGB candidates. UBVRI data~(asterisk) are plotted 
alongwith JHK~(triangle), MSX~(square) and 
IRAS data~(cross). DUSTY model fits are shown by solid lines.}
\end{figure}

\section{Notes on individual objects}

\#IRAS 17074-1845 (= Hen3-1347)\\
Henize (1976) had identified it as an H$\alpha$ emission line object.
It was classified as a hot post-AGB star on the basis of its high 
galactic latitude, far-infrared colors similar to PNe and Be spectral 
type (Parthasarathy, 1993b). Based on near-IR observations, Garc\'ia-Lario 
et al. (1997) also classified it as a post-AGB star. Parthasarathy et al. 
(2000a) classified it as a B3IIIe post-AGB star. They found 
H$\beta$ and H$\gamma$ in emission. From our H$\alpha$ image of 
this object, we did not detect 
nebulosity around the central star. The angular radius of the 
envelope inner boundary indicates a compact nebula
($<$ 2$\arcsec$).
The spectral type, dust temperature, 
double peaked spectral energy distribution and mass-loss rate 
indicate that it may be a post-AGB star evolving into the PN stage. 

\#IRAS 17395-0841 (= SS 318)\\
It was discovered as a young low excitation planetary nebula by
Vijapurkar et al. (1997). The H$\alpha$ image (300 s exposure) of the PN 
is clearly extended. Fig. 4 shows a contour plot of the object in H$\alpha$. 
The angular extent of the nebula at the FWHM of the H$\alpha$ image 
is 2.8\arcsec. It appears that bipolar outflows are just beginning to 
form in this nebula, similar to the case of Hen3-1357 (Bobrowsky et al.,
1998). High resolution H$\alpha$ images with longer exposure times may 
reveal the true morphology of this young PN. 

\begin{figure}
\epsfig{figure=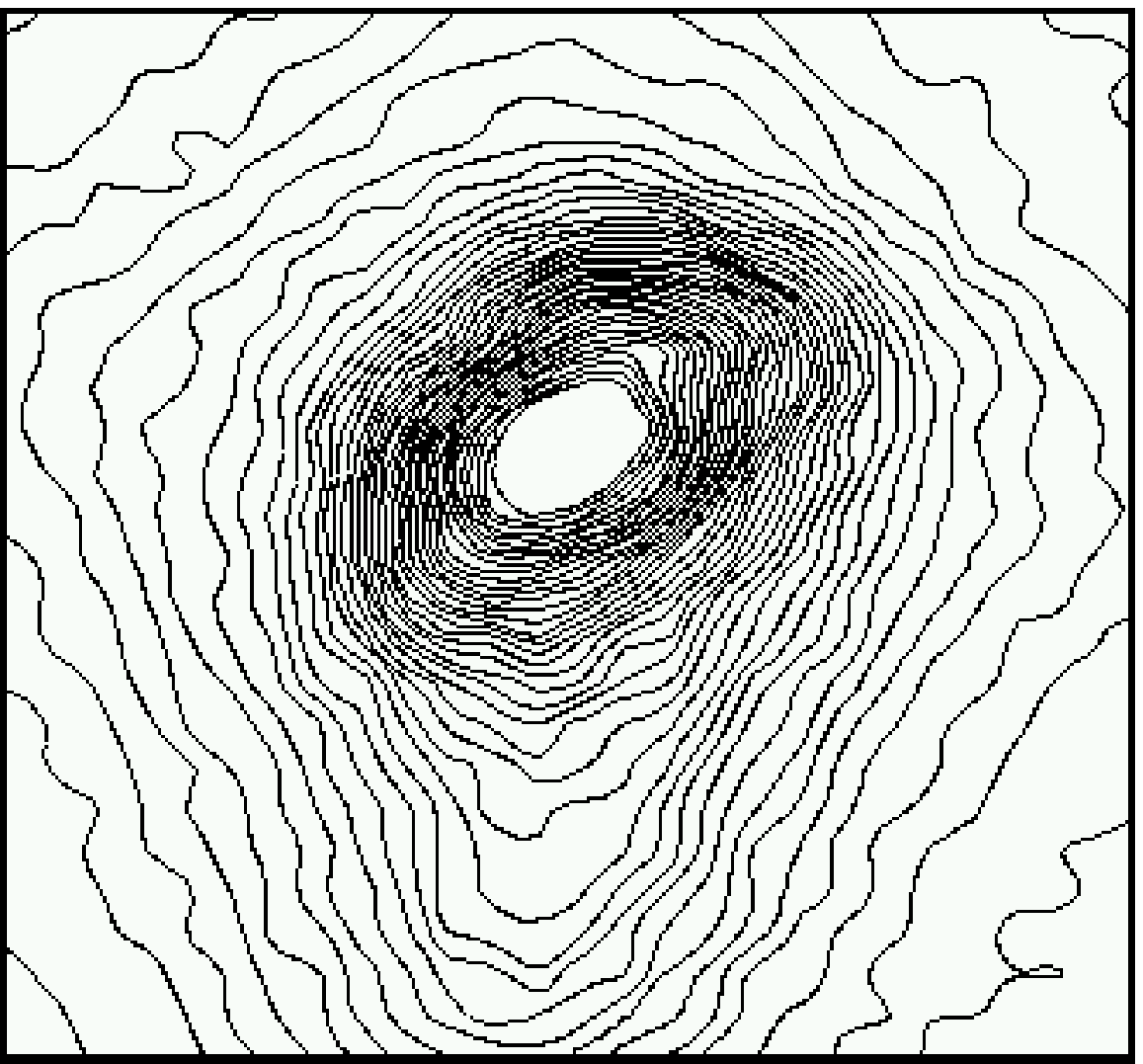,height=7.5cm,width=7.5cm}
\caption{Contour plot of the low excitation PN IRAS 17395-0841 in H$\alpha$}
\end{figure}

\#IRAS17423-1755 (= Hen3-1475)\\
On the basis of IRAS data, Parthasarathy and Pottasch (1989) first classified
it as a hot post-AGB star. It is a bipolar proto-planetary nebula (PPN)
(Bobrowsky et al., 1995, Riera et al., 1995). The collimated bipolar outflows, 
seen in our H$\alpha$ image have been studied by Borkowski et al. (1997)
and Bobrowsky et al. (1995). Wind velocities greater than 1000 km~s$^{-1}$
have been observed in this PPN (Sanchez Contreras \& Sahai, 2001,  
Borkowski \& Harrington, 2001). From our H$\alpha$ image, the south-east 
lobe appears to be more extended with an angular extent of 3.5\arcsec. 
The north-west lobe has an angular extent of only 2.2\arcsec. 
Based on [NII] HST images of the object, Borkowski et al. (1997) concluded
that Hen3-1475 is a point symmetric nebula. They found emission knots
in the outflows perpendicular to the dusty torus surrounding the central
star with the outermost pair of knots at 7.82\arcsec and 7.57\arcsec from 
the star. From the spectral energy distribution, we detected a warm dust 
component at 1500K indicating cirumstellar dust close to the central star 
as a result of ongoing post-AGB mass-loss. The dynamical age from the tip of
the AGB was estimated to be $\sim$ 4000 yrs. This value is uncertain due to the
limitations of the DUSTY code and its inability to simultaneously model the
warm and cold circumstellar dust (see Discussion below).

\#IRAS 18237-0715(= MWC930) \\
Its spectra (Parthasarathy et al., 2000a, Vijapurkar et al.,
1998) showed Balmer lines and several permitted and forbidden lines of 
[FeII] in emission. We found strong H$\alpha$ emission in this star 
indicating the presence of a low excitation PN with an angular
extent of less than 2\arcsec. Our seeing ($\sim$ 2arcsec) limited images 
could not resolve the nebula. 
There is considerable extinction (E(B$-$V)=2.1) towards the star.
At a distance of only 0.2 kpc, the heavy obscuration due to dust may be 
responsible for our inability to resolve the nebula. The effective temperature 
of the central star, spectral energy distribution and H$\alpha$ emission
suggest that it may be a hot post-AGB star. High resolution spectra and 
images of this object may help in understanding its true nature.

\#IRAS 18313-1738(= MWC939) \\
Based on strong FeII and [FeII] emission lines Allen and Swings (1976) 
classified it as a Be star. In the spectrum taken in April, 1994 with the 
1m CTIO telescope, covering the wavelength range from 3800\AA~ to 5020\AA~ 
at a resolution of 2\AA~, Vijapurkar et al.(1998), found H$\beta$, 
H$\gamma$ and several FeII and [FeII] lines in emission. Parthasarathy 
et al. (2000a) also classified it as a Be star.  We have obtained the spectrum 
of this star from 5300 to 8800 \AA with 5\AA~ pixel$^{-1}$ resolution at the 
1.02m VBO telescope (Fig. 5a). We find strong H$\alpha$ emission alongwith 
[OI], FeII, [FeII], TiII and [TiII]. The spectral features observed appear to 
be very similar to the hot central star of the proto-planetary nebula 
HD51585(Klutz \& Swings, 1977, Jascheck et al., 1996). The spectral energy 
distribution reveals the presence
of warm dust at a temperature of 1200K. The warm dust may be a signature of 
ongoing post-AGB mass-loss in this star. This object was also
detected in the A,C,D and E MSX bands. The narrow band
CCD images also revealed H$\alpha$ emission in this object. Owing to the
limitations of the DUSTY code, the warm and cold dust shells were modelled 
independent of each other and hence the derived values of r1, $\theta$ and
$\Delta$t may not be too appropriate.

We examined the IRAS low-resolution spectrum (LRS) of the object (Fig. 5b)
retrieved from http://www.iras.ucalgary.ca/iras.html 
(cf. Kwok et al., 1997).
The correction of the absolute calibration
as in Volk and Cohen (1989) and Cohen et al.(1992) has not been applied to 
the data values. While the relative shape of the LRS spectrum is OK, 
the absolute flux level need not match the IRAS 12 micron flux density 
from the photometry. 
The 9.7 $\mu$ silicate dust feature was found to be in emission. 
The 8.2, 8.6 and 11.3 $\mu$ PAH features
also appear to be in emission. The 9.7$\mu$ feature is expected to be 
in emission when the optical depth, $\tau$, of the dust envelope at 
9.7$\mu$ is less than $\sim$ 4. The peak strength of the feature occurs 
at $\tau$ $\sim$ 2. This feature originates from the circumstellar 
envelopes of oxygen-rich AGB stars. Modelling of the circumstellar dust
using DUSTY revealed the presence of graphite grains in the warm dust
close to the central star and silicate grains in the cold circumstellar dust.
The combination of C-rich and O-rich features in the circumstellar environment 
of the star may suggest a recent change to a C-rich chemistry of the outer envelope.

\setcounter{figure}{4}
\begin{figure}
\renewcommand{\thefigure}{\arabic{figure}a}
\psfig{figure=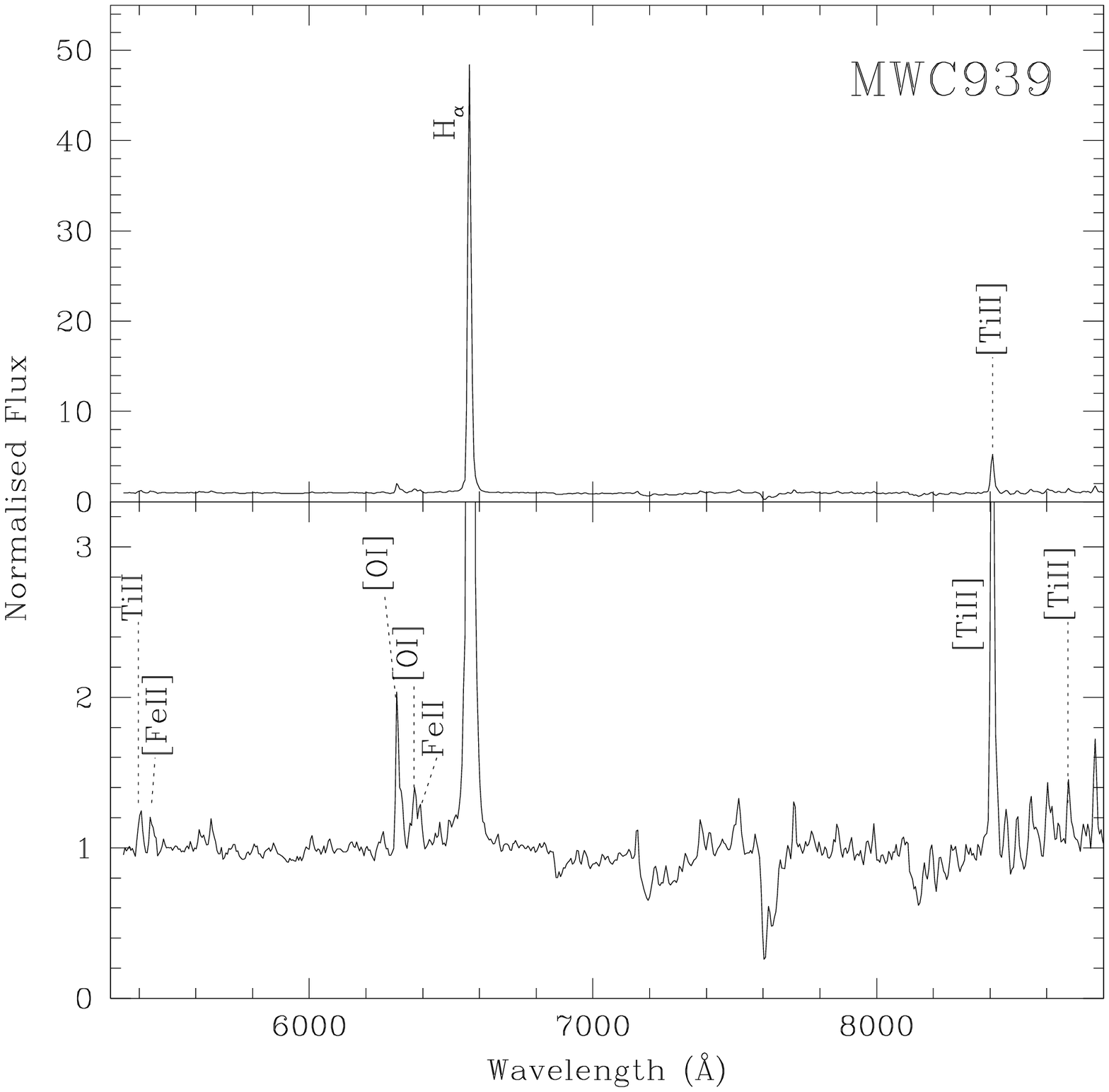,height=14cm,width=16cm}
\caption{Low resolution spectrum of the hot post-AGB candidate 
IRAS 18313-1738 (=MWC 939) }
\end{figure}

\setcounter{figure}{4}
\begin{figure}
\renewcommand{\thefigure}{\arabic{figure}b}
\psfig{figure=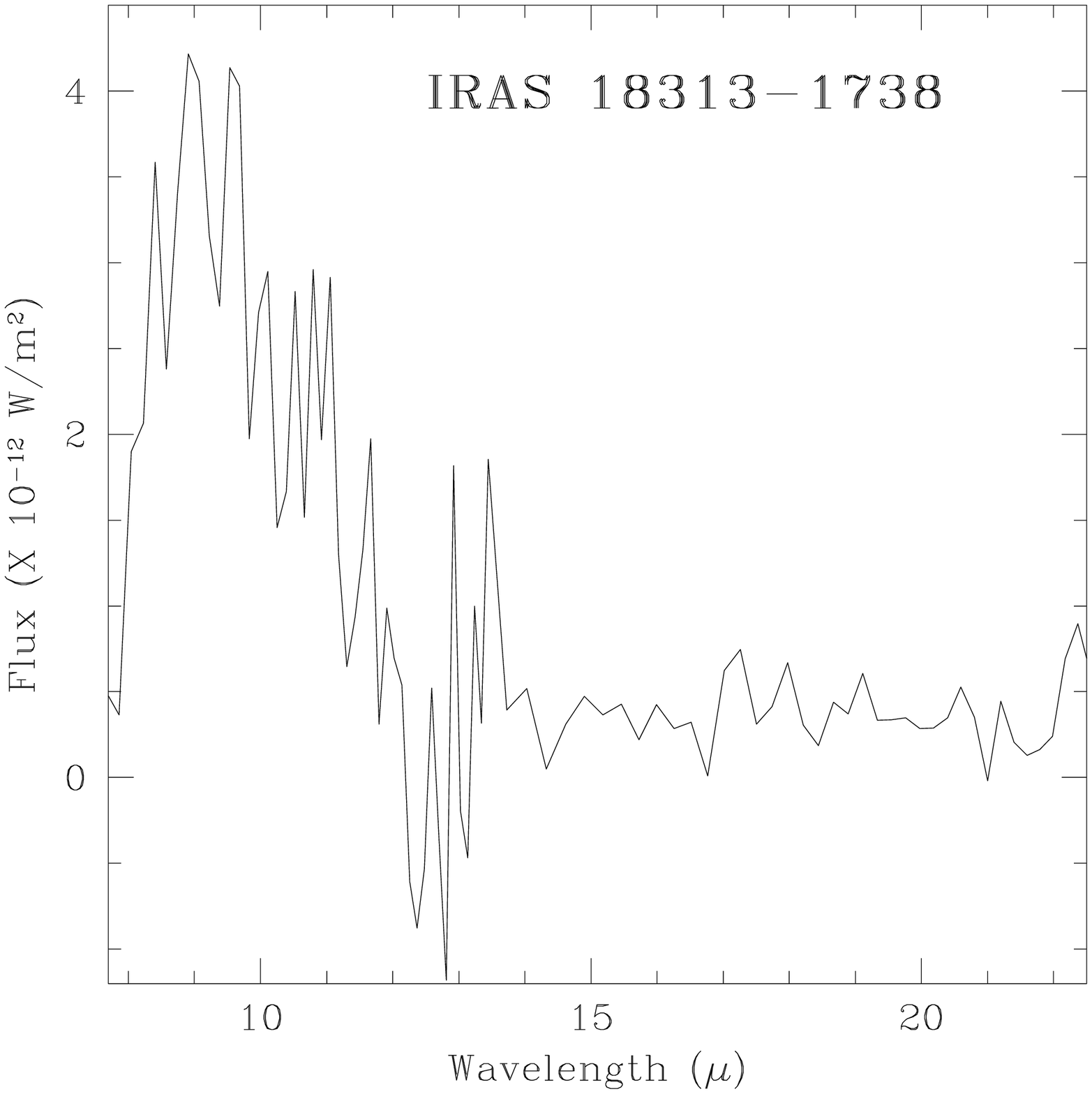,height=7.5cm, width=7.5cm}
\caption{The IRAS LRS spectrum of IRAS 18313-1738 (=MWC 939)}
\end{figure}

\#IRAS19127+1717(=SS438) \\
Whitelock \& Menzies (1986) discussed the nature of this peculiar object. 
They found that the object's spectrum has the characteristics of a 
high-density, moderate excitation PN superimposed on the continuum of a 
reddened early-type star. From the H$\beta$, H$\gamma$ and H$\delta$ lines, 
they assigned a spectral classification of B9V to the star. It is not clear 
whether the nebula is the result of mass-loss of a single star terminating 
its AGB evolution or a binary system with the outward appearance of a nebula. 
From the Balmer decrement they obtained E(B$-$V) = 1.0$\pm$0.1. Using our 
observed (B$-$V) and the intrinsic (B$-$V) for a B9V star, we 
find E(B$-$V) = 1.14. Using the DUSTY code we obtained a distance of 4.0kpc
to the central star of the PN. If on the other hand, the B9V star is the 
visual binary companion to the hot post-AGB central star of the PN, then 
using M$_{v}$=+0.2 for the B9V star, we obtain a distance of 0.2 kpc to the 
binary system. Likkel(1989) had detected OH (1667 MHz) emission in this star. 
CO emission was not detected (Likkel et al., 1991). We detected the 
presence of a warm dust component at 1300K around the star. The unusually
small estimate of the dynamical age from the tip of the AGB may be 
due to the inability to obtain a good fit to the flux distribution 
using the DUSTY code and hence the uncertainity in r1.

\#IRAS 19157-0247(= LSIV-0229) \\
This object was classified as a post-AGB star with a spectral type
of B1III by Parthasarathy et al.(2000a). From its H$\alpha$ image, we 
did not detect nebulosity around the central star. The spectral energy
distribution, high galactic latitude and  mass-loss rate suggest that it
may be a low mass star in the post-AGB phase of evolution. 

\#IRAS 19200+3457(= StHA161) \\
It was found to be an H$_{\alpha}$ emission star (Stephenson, 1986).
Preite-Martinez(1988) classified it as a possible PN. They estimated
the H$_{\beta}$ flux to be 4.8X10$^{-12}$ erg~s$^{-1}$~cm$^{-2}$ for this 
star. Using the modified Shlowskii method, they estimated a distance
of 4.3 kpc to the star. From the dust model for the star, we estimated
a distance of 3.8kpc. The angular radius of the envelope inner boundary
suggests a low excitation compact nebula ($<$ 1$\arcsec$).

\#IRAS 19399+2312(= LSII+2317) \\
It was classified as B1III by Parthasarathy et al. (2000a). Reed (1998) had 
obtained the UBV magnitudes of this star (U=11.00, B=11.20, V=10.42). We 
found them to be in good agreement with our B,V, magnitudes. The spectral
energy distribution showing the presence of a cold dust component at 127~K 
around the hot central star (T$_{eff}$ = 24000~K) and IRAS colors similar 
to PNe suggest that it may be a post-AGB star. From our short exposure 
H$\alpha$ image (120s) we did not detect nebulosity around the central star. 

\section{Discussion and Conclusions}

We have studied a small sample of hot post-AGB candidates 
selected on the basis of their IRAS colors. The near and far-infrared
flux distributions of the stars were modelled using the DUSTY code. 
We were able to estimate the inner radii of the circumstellar shells,
the distance to the stars, the mass-loss rates and angular extent of the 
inner boundary of the circumstellar envelopes from the best fit
models. We also estimated the dynamical ages from the tip
of the AGB to the present stage. However, DUSTY has certain shortcomings. 
It does not allow simultaneous modelling of warm and cold circumstellar 
dust shells. The warm and cold dust have to be modelled independent 
of each other and their simultaneous influence on the evolution of the 
nebulae cannot be accounted for.  
IRAS17423-1755, IRAS18313-1738 and IRAS19127+1717 show the presence of
both warm and cold dust in their circumstellar environment. The inner
shell radii for these three objects were calculated from the model fits to 
the cold dust shells. Since the influence of ongoing post-AGB mass-loss
(warm dust) on the remnant AGB envelope (cold dust) has not been 
accounted for, the derived r1 values in these cases are uncertain. The
uncertainity in r1 translates into an uncertainity in the dynamical ages
of the nebulae.
Independent modelling of the
warm and cold dust shells may also explain the inability to obtain a good
fit in the case of IRAS19127+1717.  Besides, DUSTY allows only a spherical
or slab geometry of the dust density distribution. This again may not
be adequate for the hot post-AGB stars rapidly evolving into PNe of 
varied morphologies. In conclusion, the physical parameters derived
from the modelling of the circumstellar envelopes are within 
the limits of the current model and may not be very precise indicators
of the circumstellar envelopes of these stars.

The dust models were generated with different grain types -
silicates, graphite and amorphous carbon. From the grain types for
the best fit models we can infer the chemical composition of the
circumstellar envelopes. IRAS17074-1845 has only silicates
in the circumstellar environment while the low excitation PN,
IRAS17395-0841 shows a carbon-rich circumstellar envelope.
Other stars such as IRAS17423-1755, IRAS18313-1738 and IRAS19127+1717
show a combination of silicates and carbon in the circumstellar
environment. The warm dust in these stars, 
is composed of amorphous carbon or graphite dust grains. 
Since the warm dust indicates ongoing post-AGB mass-loss, the
envelope chemistry appears to change from oxygen-rich to carbon-rich
as the stars evolve towards the PN stage.

The stars in this paper show Balmer lines in emission. Some 
also show several permitted and forbidden FeII lines (eg. MWC939) 
in emission and low excitation forbidden lines of SII and NII 
(eg. Hen3-1475, Riera et al., 1995). Emission lines have been detected  
in several other hot post-AGB stars (Parthasarathy et al., 2000b).
As the central stars become hotter and evolve towards the PN stage
we expect to see these emission lines in their optical spectra. 
The objects with strong emission lines in our sample are not 
young stars. These objects are not associated with star forming
regions. These are at high galactic latitudes with far-IR colors
similar to PNe. These are most likely hot post-AGB stars and as
the central stars evolve to higher temperatures they will evolve
into PNe.

\acknowledgements

We would like to thank the referee for comments and suggestions that
greatly helped in improving the manuscript.

\end{document}